# Quantifying mobile ions in perovskite-based devices with temperature-dependent capacitance measurements: frequency versus time domain

Moritz H. Futscher[1], Mahesh K. Gangishetty[2], Daniel N. Congreve[2] and Bruno Ehrler*[1]

1. AMOLF, Center for Nanophotonics, Science Park 104, 1098 XG Amsterdam, The Netherlands
2. Rowland Institute at Harvard, 100 Edwin H. Land Blvd, Cambridge, Massachusetts, United States

**Abstract**

Perovskites have proven to be a promising candidate for highly-efficient solar cells, light-emitting diodes, and X-ray detectors, overcoming limitations of inorganic semiconductors. However, they are notoriously unstable. The main reason for this instability is the migration of mobile ions through the device during operation, as they are mixed ionic-electronic conductors. Here we show how measuring the capacitance in both the frequency and the time domain can be used to study ionic dynamics within perovskite-based devices, quantifying activation energy, diffusion coefficient, sign of charge, concentration, and the length of the ionic double layer in the vicinity of the interfaces. Measuring the transient of the capacitance furthermore allows for distinguishing between ionic and electronic effects.

**Introduction**

Measuring the temperature-dependent capacitance as a function of frequency or time is a well-established technique in experimental physics to quantify electronic defect states in semiconductors.[1–4] The most famous examples are deep-level transient spectroscopy (DLTS) and thermal admittance spectroscopy (TAS).[5,6] These techniques allow for quantifying activation energy, attempt-to-escape frequency, sign, and concentration of electronic defect states.

Due to the intriguing properties of perovskites such as high charge-carrier mobilities, long diffusion lengths, strong absorption, and low exciton binding energies, perovskites have been successfully used in many optoelectronic applications including light-emitting diodes, lasers, and X-ray detectors.[7–10] Both DLTS and TAS have been used to study perovskite-based devices.[11–15] To measure electronic defect states, these techniques rely on the depletion approximation, assuming that the depletion region is free of mobile carriers. In the case of perovskites, however, this assumption is not fulfilled because they are mixed ionic-electronic conductors, leaving slow charged carriers (mobile ions) within the depletion region.[16] Consequently, temperature-dependent capacitance measurements may lead to misleading results.[17] If used correctly, measuring the capacitance as a function of frequency or time can be used to quantify these mobile ions within the perovskite bulk.[18–23]

Here we review the difference in studying mobile ions by capacitance measurements in the frequency and in the time domain. Based on the example of a PEABr$_{0.2}$Cs$_{0.4}$MA$_{0.6}$PbBr$_3$ quasi-2D/3D perovskite layer in the inverted structure consisting of NiO$_X$/perovskite/C$_{60}$/BCP, we show how both measurements can be used quantify properties of mobile ions such as activation energy, diffusion coefficient, sign of charge, concentration, and the length of the ionic double layer. Furthermore, we show how a distinction between mobile ions and electronic defect states can be made when measuring the capacitance in the time domain, allowing both mobile ions and electronic defect states to be quantified.

**Capacitance measurement**

Most optoelectronic devices such as solar cells and light-emitting diodes can be approximated as a parallel-plate device with the active material sandwiched between the two contacts. The capacitance of such a parallel-plate device is given by

$$C = \frac{\varepsilon \varepsilon_0 A}{w_D} \qquad (1)$$

where $\varepsilon_0$ is the vacuum permittivity, $A$ is the active area of the device, and $w_D$ is the depletion layer width, plus a possible additional dielectric contribution of the contact layers. In case of full depletion, $w_D$ corresponds to the thickness of the active layer, i.e the thickness of the perovskite layer, and the capacitance corresponds to the geometric capacitance of the device $C_{geo}$. The dielectric permittivity of the material $\varepsilon = \varepsilon' + i\varepsilon''$ is a complex frequency-dependent tensor. To obtain the complex capacitance of the device we measure the impedance response by applying a time-varying electric field with amplitude $V_0$ at frequency $\omega$ and measuring the current response $I_0 \sin(\omega t + \Phi)$, as indicated in Figure 1. The impedance $Z$ is defined as

$$Z(t) = \frac{V_0 \sin(\omega t)}{I_0 \sin(\omega t + \Phi)} = Z_0\, e^{-i\Phi} = Z' - iZ'' \qquad (2)$$

where $\Phi$ is the phase delay between the voltage input and the current response, and the modulus $Z_0 = V_0/I_0$ is the ratio between the amplitude of the voltage and the current signal. The complex capacitance is given by

$$C^*(\omega) = (i\omega Z)^{-1} = C'(\omega) - iC''(\omega), \qquad (3)$$

where $C'$ and $C''$ are coupled by the Kramers-Kroning relations.[24] In semiconductor devices, the measurement of the capacitance thus measures the complex dielectric permittivity of the device.

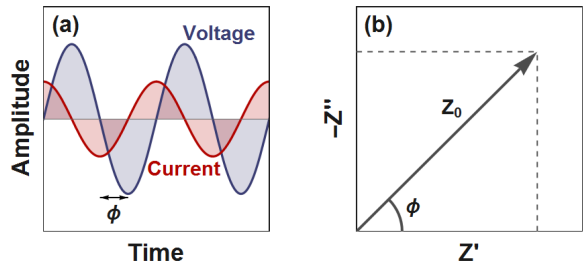

**Figure 1. (a)** Impedance measurements are based on applying a sinusoidal voltage with a certain frequency and measuring the current response. **(b)** Impedance expressed as the modulus $Z_0$ and the phase angle $\Phi$.

**Frequency versus time domain**

In a dielectric material with a Debye relaxation, i.e. a physical process exhibiting an exponential relaxation with a single time constant $\tau$, the complex dielectric response function is

$$\varepsilon(\omega) = \varepsilon'(\omega) - i\varepsilon''(\omega) = \varepsilon_\infty + \frac{\varepsilon_s - \varepsilon_\infty}{1 + i\omega\tau} \qquad (4)$$

where $\varepsilon_s$ and $\varepsilon_\infty$ are the static and asymptotic high frequency values of the dielectric constant, respectively.[25] The real part ($\varepsilon'$) is a measure of the ability to store energy in the dielectric material and the imaginary part ($\varepsilon''$) is a measure of the dielectric loss (see Figure 2a). In the imaginary part one can see a distinctive peak at a frequency of $\omega_0 = 1/\tau$, corresponding to the maximum dielectric loss in the material, i.e. the irreversible transfer of energy from the external stimulus to the dielectric material. At lower frequencies, the electrical displacement can follow the applied electric field, whereas at higher frequencies, the material has no time to respond.

Alternatively, one can instead measure the transient response of the system following an abrupt change in driving field. Using Laplace transformation to convert Equation 4 from the frequency domain into the time domain then yields

$$\varepsilon(t) = \varepsilon_\infty + (\varepsilon_s - \varepsilon_\infty)\left(1 - \exp\left(-\frac{t}{\tau}\right)\right) \quad (5)$$

which is illustrated in Figure 2b. Both the frequency and the time domain contain identical information.

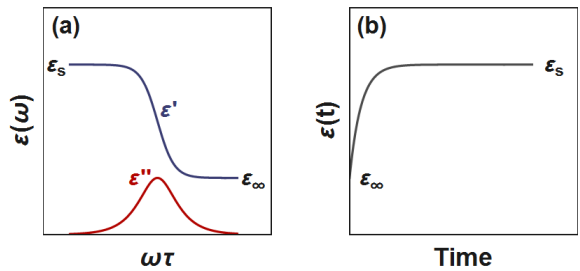

**Figure 2.** Debye dielectric material represented in **(a)** the frequency domain and **(b)** the time domain after a step function of applied electric field. Both the frequency and the time domain contain identical information.

The physical origin of a Debye process leading to the dielectric permittivity shown in Figure 2 can, for example, be the thermal emission of trapped charge carriers or the migration of mobile ions within the perovskite bulk. The magnitude of $\varepsilon_s - \varepsilon_\infty$ is then directly related to the number of trapped charge carriers or mobile ions.

In the case of electronic defect states within the band gap, the kinetics of charge-carrier capture and emission by these defect states can be described in the context of the Shockley-Read-Hall model.[26,27] In this model, the time constant measured is due to thermal emission of trapped charge carriers (see Section 1 in the Supplementary Information (SI) for details). However, the model assumes that the measured time constant results solely from the occupation/emission of electronic defect states. If a significant density of mobile ions is present within the studied material, this analysis may yield misleading results as the migration of mobile ions typically dominates the slow capacitance signal, and it hence cannot be assigned to be solely from electronic defect states.[17]

In the following we show how this dominating signal can be used to study ion migration, both in the frequency domain with impedance spectroscopy, and in the time domain with transient ion drift. Finally, we suggest that the time domain can be used to also measure electronic defect states by suppressing the signal from mobile ions.

**Impedance spectroscopy**

Impedance spectroscopy measures the capacitance as a function of frequency to observe dielectric relaxation in a material as shown in Figure 2a. To illustrate the difference between impedance spectroscopy and transient ion drift, we measure a perovskite-based device illustrated in Figure 3a (details of the device are found in Section S2 in the SI and will be published elsewhere). We measure the modulus and the phase angle at short circuit in the dark (Figure 3b) to obtain the complex capacitance from Equation 3. Figure 3c shows the real part and Figure 3d the imaginary part of the complex capacitance. At low frequencies, the phase angle is close to -90°, showing that the device operates in a capacitor-like manner. At high frequencies, the contact resistance is dominating the measured impedance response resulting in a decrease in phase angle. The real part of the capacitance shows a plateau at medium frequencies (Figure 3c). Assuming that the perovskite layer is mostly depleted at short circuit, we obtain a dielectric constant of 7.3 ± 0.1 based on the mean of three measurements, which is in good agreement with the dielectric constant obtained for $CsPbBr_3$.[28] We observe a small increase in the real part of the capacitance $C'$ at low frequencies which is accompanied with a peak in the imaginary part $C''$, similar to the dielectric relaxation shown in Figure 2a. The peak in the imaginary part shifts to lower frequencies with decreasing temperatures, which corresponds to the presence of a thermally-activated process. In perovskite-based devices such a behavior typically arises from the migration of mobile ions within the perovskite layer.[29] We hence assume in the following that the change in capacitance arises from ion migration. Later we validate this assumption with transient ion drift.

Due to the different work functions of the contacts, the perovskite layer is subject to an internal electric field in optoelectronic devices. This leads to an accumulation of mobile ions at the contact interfaces.[30] Mobile ions that accumulated at the contact interfaces form a diffuse ionic double layer that strongly influence the current injection rates of perovskites-based devices under operation.[18] The application of an oscillating voltage with a frequency $\omega$ then leads to the migration of mobile ions with a peak in the imaginary part of the complex capacitance with a maximum at the frequency

$$\omega_0 = 2\pi f_0 = \frac{1}{\tau} = \frac{D}{l^2} \qquad (6)$$

where $f$ is the measured angular frequency and $l$ the diffusion length. Often, the ion diffusion length is assumed to correspond to the Debye length $l_D$ given by

$$l_D = \sqrt{\frac{\varepsilon \varepsilon_0 k_B T}{q^2 N}} \qquad (7)$$

where $k_B$ is Boltzmann's constant, $T$ the temperature, and $N$ the doping density.[31] The diffusion coefficient can be expressed as[32,33]

$$D = \frac{\nu_0 d^2}{6} \exp\left(-\frac{\Delta G}{k_B T}\right) = \frac{\nu_0 d^2}{6} \exp\left(\frac{\Delta S}{k_B}\right) \exp\left(-\frac{\Delta H}{k_B T}\right) \qquad (8)$$

where $\nu_0$ is the attempt frequency of an ionic jump and $d$ the jump distance. $\Delta G$, $\Delta S$, and $\Delta H$ are the change in Gibbs free energy, entropy, and enthalpy during the jump of a mobile ion. The activation enthalpy is often referred to as the activation energy $E_A$. The attempt frequency is the frequency of an attempt to break or loosen a bond, related to the vibration frequency and often assumed to be in the order of $10^{12}$ s$^{-1}$.[34] Assuming that the attempt frequency is temperature independent, Equation 8 can be simplified to

$$D = D_0 \exp\left(-\frac{E_A}{k_B T}\right) \qquad (9)$$

where $D_0$ is a temperature-independent pre-factor. Assuming furthermore that the diffusion of mobile ions is primarily within the Debye layer in the vicinity to the interfaces, Equation 6 can be written as

$$\tau = \frac{\varepsilon_0 \varepsilon k_B T}{q^2 N D_0} \exp\left(\frac{E_A}{k_B T}\right). \qquad (10)$$

By plotting the measured frequency as function of temperature in an Arrhenius plot, the activation energy for ion migration can be obtained.

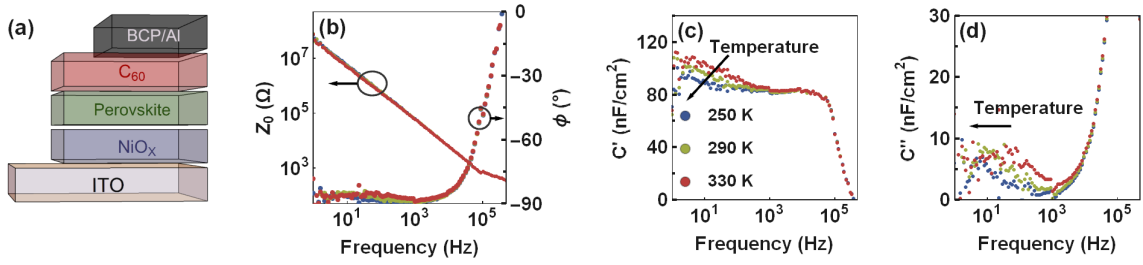

**Figure 3. (a)** Schematic representation of the device structure. **(b)** Modulus (Z) and phase angle (Φ) of the perovskite-based device measured at short circuit in the dark together with the corresponding **(c)** real and **(d)** imaginary part of the complex capacitance. The peak in the imaginary part corresponds to the temperature-dependent diffusion of mobile ions within the perovskite. The temperature dependence reveals the activation energy and the pre-factor.

Note that also in the presence of electronic defect states, an increase in capacitance at low frequencies is expected,[6] which cannot be distinguished from the effect of mobile ions. In addition, the diffusion coefficient can only be obtained if the diffusion length within the perovskite is known. Here, we assumed that the diffusion length is equal to the Debye layer. However, it has also been suggested that the migration of mobile ions extends throughout the perovskite bulk.[35,36] Furthermore, it was shown that the interfaces play an important role as charge accumulation at interfaces leads to an additional increase in capacitance.[37–40] Capacitance measurements in the time domain offer the complementary information needed to assign the observed feature to the migration of mobile ions. Only when measuring the transient of the capacitance can the diffusion coefficient be obtained without prior knowledge of the diffusion length. Transient capacitance measurements furthermore allow to distinguish between effects caused by electronic defect states and mobile ions.

**Transient ion drift**

Transient ion-drift measurements are based on measuring the capacitance transient after applying a voltage bias.[41] The voltage is chosen such that it collapses the depletion layer within the perovskite. The depletion layer width $w_D$ is approximated as[42]

$$w_D = \sqrt{\frac{2\,\varepsilon_0\,\varepsilon}{q\,N}(V_{bi} - V)} \qquad (11)$$

where and $V_{bi}$ the built-in potential. The depletion capacitance is directly related to the depletion width as

$$C_{dl} = \varepsilon_0 \varepsilon \frac{A}{w_D} = \sqrt{\frac{q\,\varepsilon_0\,\varepsilon\,N}{2\,(V_{bi}-V)}}. \qquad (12)$$

Applying a bias thus decreases the depletion layer width, increasing the measured capacitance (see Figure 4a). The capacitance of the device is obtained by an equivalent circuit model (see Section S3 in the SI for details). At higher voltages, the diffusion capacitance $C_d$ starts to dominate the measured capacitance, due to the injection of minority carriers. The diffusion capacitance is given as

$$C_d = \frac{q^2 l N_e}{k_B T} \exp\left(\frac{qV}{n k_B T}\right) \qquad (13)$$

where $N_e$ is the total equilibrium charge density at a given voltage $V$ and $n$ is the diode ideality factor.[43,44] In the case the depletion capacitance can clearly be distinguished from the diffusion capacitance, plotting $C^{-2}(V)$ allows to extract the doping density and the built-in potential of the device (see Figure 4b).[45,46] This is commonly referred to as the Mott-Schottky analysis. We obtain a built-in potential of 0.95 ± 0.05 V and a doping density of (7 ± 1) $10^{17}$ cm$^{-3}$, based on the mean of three measurements.

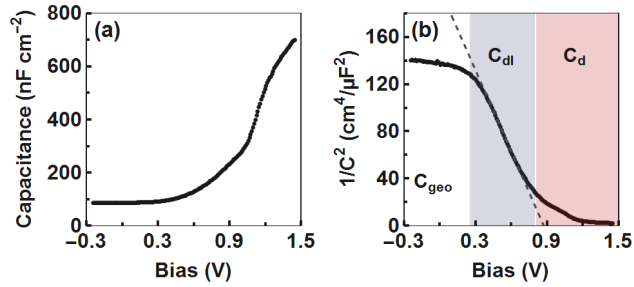

**Figure 4. (a)** Capacitance-voltage measurement and **(b)** Mott-Schottky plot of a perovskite-based device measured at 300 K in the dark at 10 kHz, illustrating different capacitive regimes. Only at a certain voltage regime can the depletion capacitance $C_{dl}$ clearly be identified. At high voltages, the diffusion capacitance $C_d$ starts to dominate the measured capacitance. The linear fit reveals the built-in potential and the doping density. The capacitance was calculated assuming a capacitor in series with a resistor (see Section S3 in the SI for details).

To measure the transient of the capacitance we apply a bias of 1.25 V, which completely collapses the depletion layer. We note that we are already in the diffusion capacitance regime when applying a 1.25 V bias voltage. However, the initial capacity change caused by the discharge due to the accumulation of minority carriers is much faster than the time resolution of our instrument. The measured capacitance rise and decay are shown in Figure 5, measured at 10 kHz where the phase angle is close to -90°, such that the measured capacitance is not affected by the series resistance of the device.

Applying a forward bias decreases the depletion width of the device. Mobile ions are now able to diffuse within the perovskite bulk which eventually leads to a uniform ion distribution within the previous depleted region. After removing the bias, the depletion width increases quickly by the movement of electric charge carriers, depleting most of the perovskite bulk according to Equation 11. Mobile ions within the depleted region will drift towards the interfaces following the internal electric field, changing the depletion width as

$$w_D(t) = \sqrt{\frac{2\,\varepsilon_0\,\varepsilon}{q\,(N \pm N_{Ion}(t))}(V_{bi} - V)} \quad (14)$$

where $N_{Ion}(t)$ is the density of mobile ions within the depletion region. For ions with the same charge as minority carriers the sign of the capacitance change is positive and for ions with the same charge as majority carriers the sign is negative. The capacitance as a function of time can thus be written as

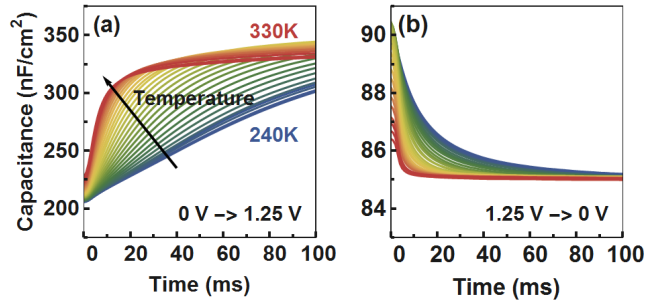

**Figure 5.** (a) Capacitance rise and (b) decay of the perovskite-based device measured at 10 kHz where the measured capacitance corresponds to the change in depletion capacitance of the device. The capacitance was calculated assuming a capacitor in parallel with a resistor.

$$C(t) = C(\infty)\left(1 \pm \frac{N_{Ion}(t)}{N}\right)^{\frac{1}{2}} \approx C(\infty)\left(1 \pm \frac{N_{Ion}(t)}{2N}\right) \quad (15)$$

where $C(\infty)$ is the junction capacitance at steady state. The approximation is valid as long as the density of the mobile ions is much lower than the doping density. To find $N_{Ion}(t)$ we assume that the thermal diffusion is negligible against drift and solve the drift equation following the work of Heiser et al.[20,41] The temporal evolution of the mobile ions drifting towards the interface can then be described by

$$\frac{\delta N_{Ion}(t)}{\delta t} = -\frac{\delta}{\delta x}N_{Ion}(t)\mu E \quad (16)$$

where $\mu$ is the mobility of mobile ions and $E$ is the electric field in the depletion region. Assuming that the electric field is static and varies linearly within the depletion region as

$$E(x) = E_0 \left(1 - \frac{x}{w_D}\right), \quad (17)$$

and that the drift of mobile ions is not affecting the electric field, we can solve Equation 16 for $N_{Ion}(t)$. Equation 15 can then be written as

$$C(t) = C(\infty) \pm \Delta C \exp\left(-\frac{t}{\tau}\right) \quad (18)$$

with the time constant $\tau$ given by

$$\tau = \frac{w_D}{\mu E}. \quad (19)$$

$\Delta C$ is the capacitance change due to the drift of mobile ions towards the interface, directly related to the mobile ion density as

$$\Delta C = C(\infty) - C(0) = C(\infty)\frac{N_{Ion}}{2N}. \quad (20)$$

By expressing the electric field as a function of the doping density as

$$E = \frac{q\, w_D N}{\varepsilon_0 \varepsilon}, \quad (21)$$

together with the Einstein relation, the time constant can be written as

$$\tau = \frac{k_B T \varepsilon_0 \varepsilon}{q^2 DN} \quad (22)$$

with the diffusion coefficient given by Equation 9. This equation can now be used to extract the diffusion coefficient and activation energy from the measured capacitance transients. Note that this equation is the same as Equation 10, which describes the time constant for ion migration in the frequency domain.

Note that we have assumed a linear electric field within the depletion region and that the electric field is unaffected by the drift of mobile ions, which is only true if the density of mobile ions is small compared to the background doping density. In the case of a mobile ion density close to the background doping density, the transient would have a non-exponential behavior making the analysis more complex.[20] We have furthermore assumed that the total ion concentration is conserved, i.e. mobile ions are not diffusing into the contact layers.

To distinguish mobile ions from electronic defect states we use the difference in rise and decay time of the capacitance change. In the case of mobile ions, the rise time (Figure 5a) due to diffusion of mobile ions is expected to be longer than the decay time due to drift of mobile ions (Figure 5b). In contrast, the capture-rate for electronic defect states is much higher than the emission rate, resulting in a much faster rise time compared to the decay time. Measuring the rise and the decay of the capacitance thus allows to distinguish between capacitance changes due to electronic defect states and mobile ions (see Section S4 in the SI for details).

A unique feature of transient ion-drift measurements is that one can differentiate between mobile cations and mobile anions. Here, the negative capacitance change in Figure 5b measures an increase in depletion width due to ions migrating towards the contact interfaces. The Mott-Schottky plot shows a p-type behavior of the perovskite (Figure 4), the negative capacitance change in Figure 5b hence corresponds to the migration of an anion. We thus attribute the measured changes in capacitance due to bromide migration, presumably due to vacancy-mediated migration.[47,48]

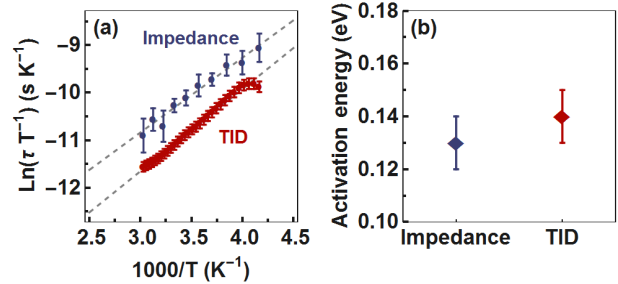

**Figure 6.** Arrhenius plot of the measured thermal emission rates by **(a)** impedance spectroscopy and **(b)** transient ion drift (TID). **(c)** Activation energies obtained from the two measurement techniques based on the mean out of three measurements each.

The Arrhenius plots together with the obtained activation energies for both impedance spectroscopy and transient ion-drift measurements are shown in Figure 6. For impedance spectroscopy and transient ion-drift measurements we obtain very similar activation energies of 0.13 ± 0.01 and 0.14 ± 0.01 eV, respectively. The obtained activation energy is close to theoretical predictions and experimental observations for the migration of bromide in MAPbBr$_3$ and CsPbBr$_3$ (0.09 to 0.25 eV).[49–51]

From the capacitance transients we obtain a concentration of (5.1 ± 2.5) x 10$^{16}$ cm$^{-3}$ and a diffusion coefficient of (3.1 ± 0.4) x 10$^{-11}$ cm$^2$/s for the mobile anions. We can now use this coefficient together with the frequency of the dielectric loss peak (Figure 3d) to obtain a diffusion length for ion migration of 6.2 ± 0.4 nm from Equation 6. The calculated diffusion length is larger than the Debye length of 3.9 ± 0.3 nm calculated from Equation 7. Hence, the assumption that the diffusion of mobile ions during frequency-dependent capacitance measurements is limited to the Debye layer is not fulfilled. This underestimation of the diffusion length explains the difference between impedance spectroscopy and transient ion-drift measurements in Figure 6a. Note that we assume, for the purpose of the comparison between transient ion drift and impedance measurements, that the ion migration is the same near the interface and through the bulk.

**Conclusion**

Capacitance techniques must be applied with caution to mixed ionic-electronic conductors, because the measured capacitance features can be caused by mobile ions rather or electronic defect states. We have shown that transient ion-drift measurements have the virtue that a distinction between electronic defect states and mobile ions can be made. In addition, transient ion drift allows fast and non-destructive quantification of activation energy, diffusion coefficient, sign of charge, and concentration of mobile ions in perovskite-based devices. Using the diffusion coefficient obtained by transient ion drift, the length of the ionic double layer can be determined from impedance spectroscopy measurements.

Since the migration of mobile ions is a key degradation mechanism in perovskite-based devices, reducing ion migration is crucial for the fabrication of stable devices. Capacitance techniques provide a tool to systematically investigate the effects of different passivation agents, fabrication methods, and perovskite compositions on ion migration in full perovskite-based devices guiding the way to long-lasting devices, critical for commercialisation.

**Acknowledgements**

The authors thank Esther Alarcón Lladó for carefully reading and commenting on the manuscript. This work is part of the research program of the Netherlands Organization for Scientific Research (NWO). D.N.C. and M.K.G. acknowledge the support of the Rowland Fellowship at the Rowland Institute at Harvard University.

# SUPPLEMENTARY INFORMATION FOR

# Quantifying mobile ions in perovskite-based devices with temperature-dependent capacitance measurements: frequency versus time domain

*Moritz H. Futscher[1], Mahesh K. Gangishetty[2], Daniel N. Congreve[2] and Bruno Ehrler*[1]*

1. AMOLF, Center for Nanophotonics, Science Park 104,

   1098 XG Amsterdam, The Netherlands

2. Rowland Institute at Harvard, 100 Edwin H. Land Blvd,

   Cambridge, Massachusetts, United States

**Corresponding Author**

* ehrler@amolf.nl

## S1 Shockley-Read-Hall model

In the Shockley-Read-Hall model, the thermal emission rate of trapped electrons is given by

$$\frac{1}{\tau} = e_n = \sigma_n v_{th} N_C \exp\left(-\frac{E_T}{k_B T}\right)$$

where $\sigma_n$ is the capture cross section for capturing electrons, $v_{th}$ the thermal velocity, $N_C$ the density of states in the conduction band, $E_T$ the depth of the trap from the conduction band, $k_B$ Boltzmann's constant, and $T$ the temperature.[1,2] Taking the temperature dependence of the pre-factors into account, this can be simplified to

$$e_n = A(T) \exp\left(-\frac{E_T}{k_B T}\right)$$

where $A = A_0 T^2$ is the attempt-to-escape frequency. An analogous expression holds for the thermal emission rate of trapped holes. By measuring the time constant as a function of temperature, the depth of the trap and the attempt-to-escape frequency can be obtained. When the time constant is measured from the transient of the capacitance, this technique is called deep-level transient spectroscopy (DLTS) originally developed by Lang et al.,[3] and when the time constant is measured from the frequency-dependence of the capacitance, this technique is called thermal admittance spectroscopy (TAS) originally developed by Walter et al.[4]

## S2 Experimental

**Device fabrication.** ITO substrates were cleaned by sonicating sequentially in detergent (Micron - 90), water (twice) and acetone (twice) for 10 minutes each, before soaking them in boiling isopropanol for 20 minutes to remove the leftover organic contamination. The substrates were then treated with $O_2$ plasma for 5 minutes using 0.5 Torr $O_2$ gas at 500 W. On top of these cleaned substrates, 30 µL precursors solution of $NiO_X$ was spin coated at 2500 rpm with an acceleration of 2500 rpm/s. The films were then immediately transferred to a hot plate, which was kept at 100 °C and annealed at 300 °C for 3 hours. After cooling the $NiO_X$ films, a thin layer of perovskite was made by spin coating the perovskite precursor solution at 1000 rpm for 10 seconds and ramped up to 3000 rpm for 45 seconds. After 20 seconds, 90 µL of chloroform (anti-solvent) was dripped on the spinning perovskite layer. The perovskite precursor was prepared by mixing appropriate ratios of 0.3 M of MABr, CsBr, $PbBr_3$ and PEABr to get $PEABr_{0.2}Cs_{0.4}MA_{0.6}PbBr_3$ in DMF/DMSO. The detailed preparation of precursor solution is reported in our previous work.[5,6] After coating the perovskite layer, the films were taken into an evaporation chamber where, 50 nm $C_{60}$, 5 nm BCP, and 60 nm Ag were sequentially deposited at $10^{-6}$ mbar.

**Electrical measurements.** All measurements were performed in the dark in a Janis VPF-100 liquid nitrogen cryostat at pressures below 2 x $10^{-6}$ mbar. Samples were loaded into the cryostat inside a nitrogen-filled glovebox to avoid air exposure. Capacitance-voltage characteristics were measured at 10 kHz with an AC frequency of 50 mV from 1.45 to -0.25 V with a step size of 0.01 V. Impedance spectroscopy measurements were performed with an AC voltage of 50 mV at short circuit between 240 and 330 K in steps of 10 K. Transient ion-drift measurements were performed at 10 kHz with an AC voltage of 20 mV between 240 and 330 K in steps of 2 K. The transients were averaged over 20 separate measurements. The temperature accuracy was set to 0.2 K.

**S3 Equivalent circuit model**

The complex impedance of a parallel-plate device can be modelled with a resistor in parallel and a resistor in series to a capacitor, as shown in Figure S1.

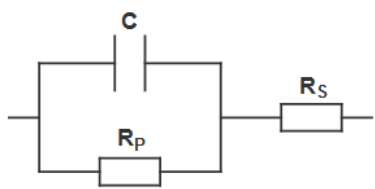

**Figure S1.** Circuit model of a resistor in parallel ($R_P$) and a resistor in series ($R_S$) with a capacitor (C).

If the capacitance is measured at one frequency, however, the complex impedance must be obtained with a model consisting only of two free parameters, i.e. one resistor and one capacitor. Time-dependent and voltage-dependent capacitance measurements can thus be modelled with a resistor either in parallel or in series to the capacitor as shown in Figure S2 and Table S1. At high impedance the series resistance can be neglected, at low impedance the shunt resistance can be neglected. As long as the device does not suffer from resistive losses, however, the capacitance is identical for both models.

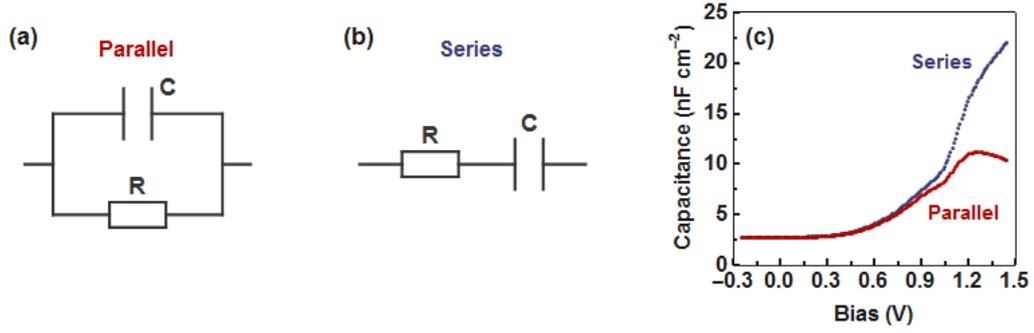

**Figure S2.** Circuit model of a capacitor (C) **(a)** in parallel and **(b)** in series with a resistor (R). **(c)** Capacitance as a function of voltage of a perovskite-based device calculated assuming a capacitor in series or in parallel with a resistor measured at 300 K in the dark at 10 kHz.

To measure the transient of the capacitance, we neglect the series resistance assuming a resistor in parallel with a capacitor. We thus choose the measuring frequency such that the impedance is not limited by the series resistance of the device, i.e. the phase angle has to be close to -90°.

**Table S1.** Comparison of capacitance calculated with a capacitor either in series or in parallel with a resistor.

|  | Series | Parallel |
|---|---|---|
| **Impedance** | $R + \dfrac{1}{i\omega C}$ | $\left(R + \dfrac{1}{i\omega C}\right)^{-1}$ |
| **Capacitance** | $-\dfrac{1}{\omega\, \mathrm{Im}(Z)}$ | $\dfrac{1}{\omega}\mathrm{Im}\left(\dfrac{1}{Z}\right)$ |

Due to charge-carrier injection at forward biases, the series resistance can often no longer be neglected. In this case, the capacitance can be calculated as a function of voltage assuming a capacitor in series with a resistor. Figure S2c shows the difference in capacitance as a function of voltage assuming a capacitor in series or in parallel with a resistor. At low biases, the difference between the two models is negligible. At high biases, there is a growing difference in capacitance obtained from the two models. In our case, however, the differences between the two models in the values obtained with the Mott-Schottky analysis are small, i.e. 0.89 and 0.87 V for the built-in potential and 8.0 x $10^{17}$ and 8.1 x $10^{17}$ cm$^{-3}$ for the doping density, for a capacitor in parallel and in series with a resistor, respectively.

**S4 Rise and decay time**

The characteristic rise and decay times for both mobile ions and electronic defect states are shown in Table S2. For mobile ions, the rise time is assumed to be due to diffusion from the interfaces into the perovskite bulk. The decay time is assumed to be due to drift of mobile ions from the bulk towards the interfaces, i.e. $\tau = \frac{w_D}{\mu E}$, where $w_D$ is the depletion width and $\mu$ is the mobility of mobile ions. The electric field is approximated as $E = \frac{V}{w_D}$. The rise and decay time follow from the Shockley-Read-Hall model. Consequently, electronic defect states and mobile ions show different ratios between rise and decay time, so that a distinction can be made between them.

**Table S2.** Comparison of rise and decay times of electronic defect states and mobile ions at room temperature assuming a defect state with a trap energy between 0.2 and 0.6 eV. $D$ is the diffusion coefficient of mobile ions, $q$ the elementary charge, and $V_{bi}$ the built-in potential.

|  | Electronic defect states | Mobile ions |
|---|---|---|
| **Rise time** | $\frac{1}{\sigma v_{th} N} \approx 10^{-13} - 10^{-10}$ s | $\frac{w_D^2}{D} \approx 10^{-1} - 10^1$ s |
| **Decay time** | $\frac{1}{\sigma v_{th} N} \exp\left(\frac{E_T}{k_B T}\right) \approx 10^{-10} - 10^0$ s | $\frac{w_D^2 k_B T}{D q V_{bi}} \approx 10^{-3} - 10^0$ s |

Figure S3 shows measured capacitance transients after a filling voltage of 1.25 V for different periods of time. Applying a voltage pulse for 2 seconds allows mobile ions to be measured, since mobile ions have sufficient time to diffuse from the interfaces into the perovskite bulk (Figure S3a). On the other hand, applying a voltage bias for 0.2 milliseconds only fills electronic defect states present in the material, as ions have not enough time to react (Figure S3c). When applying a bias for 20 milliseconds both ions and electronic defects states are measured (Figure S3b). The Arrhenius plot of the measured thermal emission rates for electronic defect states is shown in Figure S3d. We obtain an activation energy of 0.08 ± 0.01 eV, a concentration of 3 x $10^{15}$ cm$^{-3}$ and an attempt-to-escape frequency of 1.3 ± 0.4 $10^{13}$ s$^{-1}$. The obtained concentration of electronic defect states is much lower than the concentration of mobile ions. We therefore assume that the presence of electronic defect states is not influencing the our measurements of mobile ions.

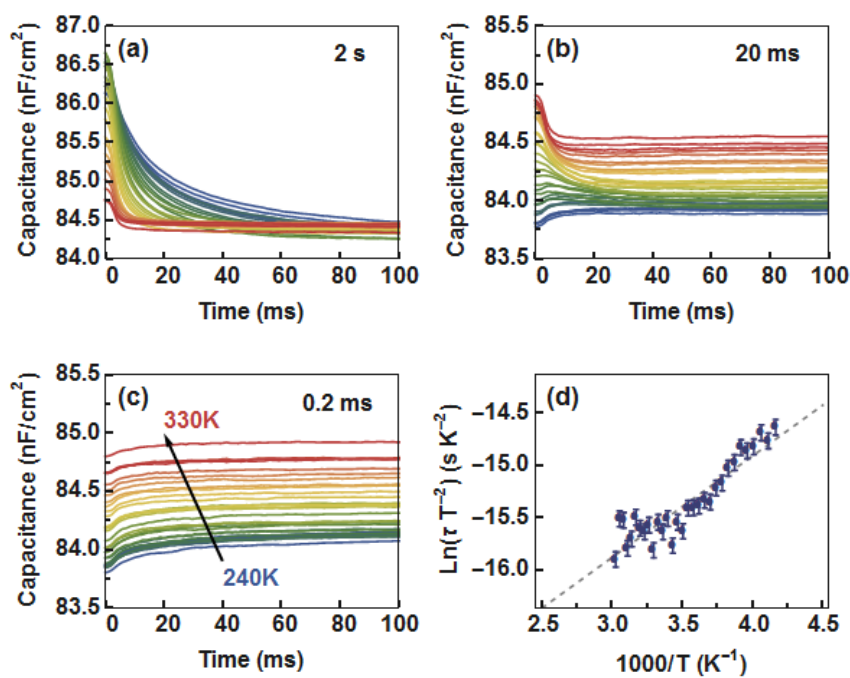

**Figure S3.** Capacitance transient measurements between 240 and 330 K in steps of 3 K measured at 0 V at 10 kHz. Capacitance transients are measured after applying a bias of 1.25 V for **(a)** 2 seconds, **(b)** 20 milliseconds, and **(c)** 0.2 milliseconds. **(d)** Arrhenius plot of the thermal emission rates from (c). The linear fit reveals the activation energy and the attempt-to-escape frequency of electronic defect states.